\documentclass{aa}  

\usepackage{comment}

\usepackage{threeparttable}

\usepackage{graphicx}
\usepackage{txfonts}
\usepackage{lipsum}
\usepackage{subcaption}         % necessary for continued figures, example in section 3
\usepackage{lscape}             % to rotate a single page table, example in appendix.
\usepackage{placeins}           % useful with \FloatBarrier, to keep 
                                \usepackage{hyperref}
\hypersetup{
    colorlinks=true,
    linkcolor=blue,
    filecolor=magenta,      
    urlcolor=blue,
    citecolor=blue,
    } 
\begin{document}

   \title{Search for radio polarization in the particle-accelerating colliding-wind binaries WR~147 and HD~167971}

   \author{A. B. Blanco\inst{1}, 
   M. De Becker\inst{1},
   P. Benaglia\inst{2},
        \and S. del Palacio\inst{3}
        }

   \institute{Space Sciences, Technologies and Astrophysics Research (STAR) Institute, University of Liège, Quartier Agora, 19c,
Allée du 6 Août, B5c, 4000 Sart Tilman,
               Belgium \\
             \email{ablanco@uliege.be}
            \and Instituto Argentino de Radioastronomia (CONICET; CICPBA; UNLP), C.C. No 5, 1894, Villa Elisa, Argentina
            \and
            Department of Space, Earth and Environment, Chalmers University of Technology, 412 96 Gothenburg, Sweden
            \\}

   \date{Received 23 February 2026 / Accepted 20 April 2026}

% \abstract{}{}{}{}{}
% 5 {} token are mandatory
 
  \abstract
   {Particle-accelerating colliding-wind binaries (PACWBs) are multiple systems of massive stars in which strong stellar winds collide, accelerating particles to relativistic energies. This population of relativistic particles emits NT radiation, including synchrotron radiation in the radio domain. This emission is expected to be linearly polarized, but the polarization signature has not yet been detected for a PACWB.}
   {Our objective is to quantify the linear polarization of synchrotron radiation in two well-known PACWBs and to interpret our measurements within the framework of the physics of these specific NT emitters.} 
   {We observed the PACWBs WR~147 and HD~167971 with the Very Large Array (VLA) radio interferometer in the frequency bands L and C (1--2 and 4--8~GHz, respectively), where synchrotron emission is expected to be more prominent. We performed polarization calibration and analyzed the resulting Stokes maps.}
   {We did not detect any polarization signature for either of the two targets in either of the two bands, even when considering narrower bands to mitigate the effect of bandpass depolarization. The most conservative upper limit on the polarization degree is on the order of 1\,\% for both targets. }
   {The lack of linear polarization for the two targets is likely attributable to a combination of effects, including the turbulent nature of the magnetic field in the synchrotron-emitting region, and depolarization processes based on Faraday rotation that are certainly active in these sources. Their complex geometry, unresolved by the VLA at these frequencies, is most likely to lead to beam depolarization. We emphasize that, in contrast to other canonical synchrotron sources, PACWBs are also subject to thermal dilution. This is especially relevant for systems with stars whose winds are strong enough to contribute copiously to thermal emission, such as those harboring a Wolf-Rayet component.}


\titlerunning{Radio polarization in PACWBs}
\authorrunning{A. B. Blanco}

   \maketitle

   \nolinenumbers

\section{Introduction}

High-mass stars ($M \ge 8~\mathrm{M_\odot}$) are a type of hot, luminous star commonly found gravitationally bound in binary or multiple systems, with a binary fraction of at least 90~$\%$ for O-type stars in young clusters \citep{Offner2023}. They are characterized by strong, fast, and dense winds of ejected material that cause the environments of these stars to become irreversibly modified. The outflowing material that is expelled from the massive-star surface in the winds can reach terminal velocities ($v_{\infty}$) of a few 1000~~$\mathrm{km\,s}^{-1}$, with mass-loss rates ($\dot{M}$) that typically range between 10$^{-7}$ and 10$^{-5}$~$\mathrm{M_{\odot}\,yr}$$^{-1}$. These wind parameters are intimately dependent on the spectral type and luminosity class of the star. In particular, Wolf-Rayet (WR) stars, which are the evolved counterparts of O-type stars, are typically associated with significantly stronger winds.

When a multiple stellar system is made of two or more high-mass stars, their powerful winds collide, resulting in what is known as a colliding-wind binary (CWB). Within the category of CWBs, some systems show evidence of being capable of accelerating charged particles, such as electrons and protons, to relativistic speeds. We refer to these systems as particle-accelerating colliding-wind binaries (PACWBs, \citealt{DeBecker2013, DeBecker2017}). Thus far, more than 50 objects have been catalogued as PACWBs \citep{DeBecker2013}\footnote{The catalogue is regularly updated at \url{https://www.astro.uliege.be/~debecker/pacwb/}.}. The process responsible for particle acceleration in PACWBs is believed to be diffusive shock acceleration (DSA, \citealt{Drury1983D}). DSA occurs in the shocks produced by the collision of winds in the colliding-wind region (CWR) and allows particles to gain energy through iterative interactions. These relativistic particles will, in turn, emit nonthermal (NT) radiation through different radiative processes across the electromagnetic spectrum \citep{EU1993, Pittard2006, Pittard2006b, Reimer2006, DeBecker2007, Pittard2021}. In particular, relativistic electrons in the presence of a magnetic field end up producing NT synchrotron radio emission. In PACWBs, the magnetic field present in the CWR, which is of stellar origin, interacts with the relativistic electrons accelerated in the shocks, and as a result, synchrotron radiation is produced. 

Synchrotron radio emission is believed to be highly linearly polarized. Theoretical values for the polarization degree, $\Pi$, defined as the fraction of the total intensity that is intrinsically linearly polarized, are expected to be at the level of about $70~\%$. Polarized synchrotron radiation has, for example, been measured and well characterized in supernova remnants (SNRs), where particle acceleration and NT emission processes are known to be active \citep{Dubner2015, Cotton2024}. However, such a polarization signature has not yet been measured in the synchrotron emission of a PACWB, with only one attempt at detection reported in the literature thus far \citep{Hales2017}. Determining $\Pi$ for a synchrotron source could help us infer the degree of turbulence and the characteristics of the magnetic field in the CWR. In this sense, our goal is to investigate the question of the polarization of the synchrotron radiation in PACWBs and to improve our knowledge of the NT emission in the environment of PACWBs. To this end, we started by selecting two well-known PACWBs known to be very bright at radio wavelengths, WR~147 and HD~167971. We observed them in the ranges of the radio domain where synchrotron emission is stronger (i.e., at GHz frequencies), where this emission is significantly produced and not too severely attenuated by free-free absorption (FFA). 

We introduce the sample studied in Sect.\,\ref{sample}. We describe the observations and data reduction process in Sect.\,\ref{obs}. The results are presented in Sect.\,\ref{results}. In Sect.\,\ref{disc}, we debate the physical interpretation of the results. A set of conclusions derived from the analysis is given in Sect.\,\ref{concl}.

\section{Selected targets}\label{sample}

\subsection{Sample description}

In this work, we examine two massive star systems: one that includes a WR star and another harboring O-type stars only. Our approach aims at complementing the findings of \citet{Hales2017}, while covering a wider region of the parameter space for this class of objects.

The massive system WR~147 is located in the Cygnus region and consists of a WN8 star and an early B-type or O8-9~V-III companion. These two stellar components are separated by a projected angular distance of approximately 0$\farcs$6, as measured from VLA radio observations \citep{Churchwell1992}. For an adopted distance of 1.7~kpc \citep{Crowther2023}, this corresponds to a projected linear separation of about 1000~AU. Such a separation implies a very long orbital period, which has not yet been determined, but is thought to be on the order of a thousand years. Despite the large physical separation between the two stellar components, the system remains unresolved in our observations (see the angular resolutions achieved in Table~\ref{StokesI-Details}, Sect.~\ref{results}). WR~147 is known to be one of the most luminous synchrotron emitters among PACWBs, along with WR~146, which was the target of the study by \citet{Hales2017}.

HD~167971 is a member of the NGC~6604 open cluster located at a distance of about 1.7~kpc \citep{Reipurth2008}. It consists of a hierarchical triple system made of a close binary (O6-7 V + O6-7 V) with an orbital period of 3.3 days and an O8 supergiant that orbits the inner binary with a period of about 21~years \citep{DeBecker2012, Ibanoglu2013, LeBouquin2017}. The outer orbit of the system has been resolved using near-infrared (NIR) long-baseline interferometry at the VLTI, yielding a projected angular semi-major axis $a$ of about 18~mas, with the separation varying along the orbit due to its significant eccentricity \citep{LeBouquin2017}. At a distance of 1.7~kpc, this is correlated with a projected linear separation of roughly 30–45~AU. The combined winds of the inner binary collide with the wind of the supergiant, producing the population of relativistic particles that, in turn, are responsible for the synchrotron radiation emission. HD~167971 is also known as the brightest synchrotron radio emitter among O-type PACWBs.

\subsection{Previous radio measurements}

Both targets have been thoroughly studied in the radio domain in the past. However, no attempts have been reported to measure the polarization degree of their synchrotron radiation. The synchrotron emission region has been resolved due to high-angular resolution observations for both WR~147 \citep{Churchwell1992, Williams1997} and HD~167971 \citep{SanchezBermudez2019, DeBecker2024}. In the range of frequencies of interest to this work, where synchrotron emission is stronger, and the thermal dilution caused by the winds is not too severe, both targets benefit from previous observations. 

WR~147 has been observed in the MHz range with the Westerbork Synthesis Radio Telescope (WSRT) and the Giant Metrewave Radio Telescope (GMRT) by \citet{SetiaGunawan2003} and \citet{Benaglia2020}, respectively, and flux measurements have been reported. The system has also been studied at centimeter wavelengths using Very Large Array (VLA) data \citep{Skinner1999, Tasseroul2025}. In particular, the study by \citet{Tasseroul2025} presented a detailed account of the radio spectral energy distribution of WR~147. This source is quite unique from the point of view of both the strength of the thermal emission from the WR wind and the brightness of its synchrotron component.

HD~167971 has also been studied extensively at centimetric wavelengths, mainly using VLA data. In addition to sporadic flux measurements obtained a few decades ago \citep{Bieging1989, Phillips1990, Contreras1996}, \citet{Blomme2007} produced light curves at several wavelengths, allowing us to obtain an overview of the radio behavior of the system over its long orbit. The radio light curves clearly indicate significant variability correlated with the orbital phase, with a maximum that coincides with the system's periastron passage.

\section{Observations and data reduction}\label{obs}

The targets WR~147 and HD~167971 were observed as part of an observational campaign with the Karl G. Jansky Very Large Array (VLA) radio interferometer, located in New Mexico, USA. The observations took place in August 2016, while the array was in configuration B. According to the ephemeris published by \citet{LeBouquin2017} for HD~167971, these dates correspond to an orbital phase that lies in the 0.34--0.39 range (taking into account the uncertainty on the period, i.e., 8706 $\pm$ 540\,d). This corresponds to an orbital phase where the synchrotron emission is measured significantly away from periastron, thus preventing any substantial FFA from reducing the measured flux. For WR~147, the lack of published orbital elements prevents us from speculating on the orbital phase.

We performed full polarization observations in two frequency bands: the L band (20~cm, centered at 1.5~GHz) and C band (5~cm, centered at 6~GHz), spanning bandwidths of 1~GHz and 4~GHz, respectively. The correlator was configured to deliver 8 spectral windows in the 20~cm band, and 32 windows in the 5~cm band, with 64 channels in each window. With each channel spanning 2~MHz, the total bandwidth per spectral window was 128~MHz. The total observing time on-source was 36~min~36~s at the L band and 13~min~6~s at the C band for WR~147, and 37~min~57~s at the L band and 13~min~6~s at the C band for HD~167971. The calibrator 3C~286 was observed for flux density, bandpass, and polarization position angle calibration. QSO B2005+403 and QSO J1745-0753 were used for amplitude and phase calibration for WR~147 and HD~167971, respectively, and the unpolarized source J2355+4950 was observed for instrumental leakage calibration. 
Data reduction was performed using \texttt{CASA} \citep[Common Astronomy Software Applications;][]{McMullin2007}. This included the flagging of bad data and data affected by radio frequency interference (RFI), followed by the standard gain and bandpass calibration. Polarization calibration was carried out using standard \texttt{CASA} polarization calibration routines, following the National Radio Astronomy Observatory (NRAO) guidelines. 
We solved for the cross-hand delays (with a multiband delay approach), followed by the determination of the instrumental polarization leakage terms (D-terms) and the absolute polarization position angle using the task \texttt{polcal}. The derived calibration solutions were then applied prior to imaging the targets in full Stokes parameters. The final products include Stokes $I, Q$, and $U$ maps (see Table~\ref{StokesI-Details}, Sect.~\ref{results}, for the angular resolution and sensitivities achieved). Given that the targets were not expected to be spatially resolved at the angular resolutions reached by the VLA in these bands, for the imaging we adopted Briggs weighting (with robust = 0.5) to favor sensitivity without significantly degrading the angular resolution.

\section{Results}\label{results}

\begin{figure*}[t]
\centering
\setlength{\tabcolsep}{2pt}

\begin{tabular}{ccc}
\textbf{Stokes I} & \textbf{Stokes Q} & \textbf{Stokes U} \\[4pt]
\includegraphics[width=0.34\textwidth]{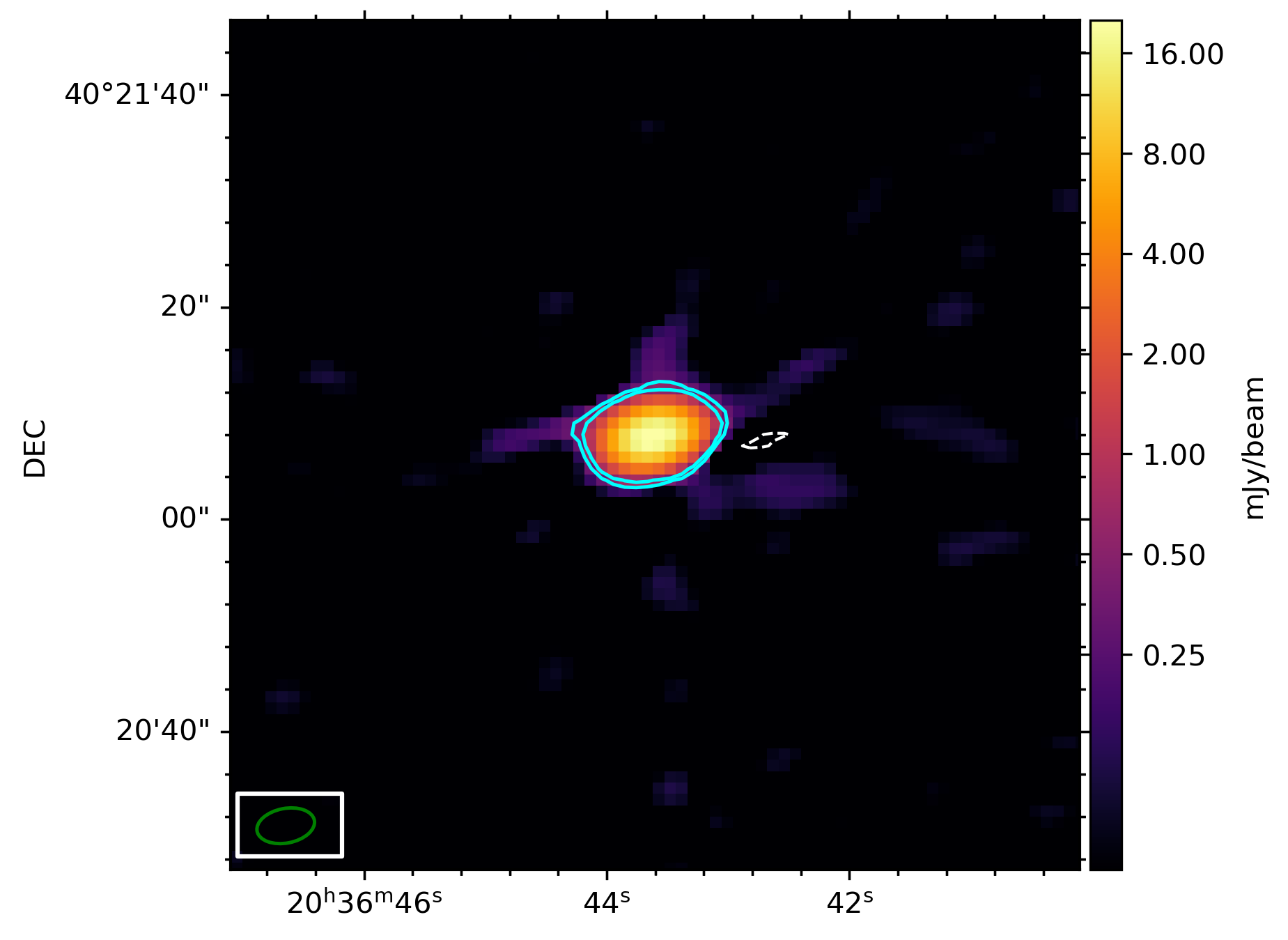} &
\includegraphics[width=0.33\textwidth]{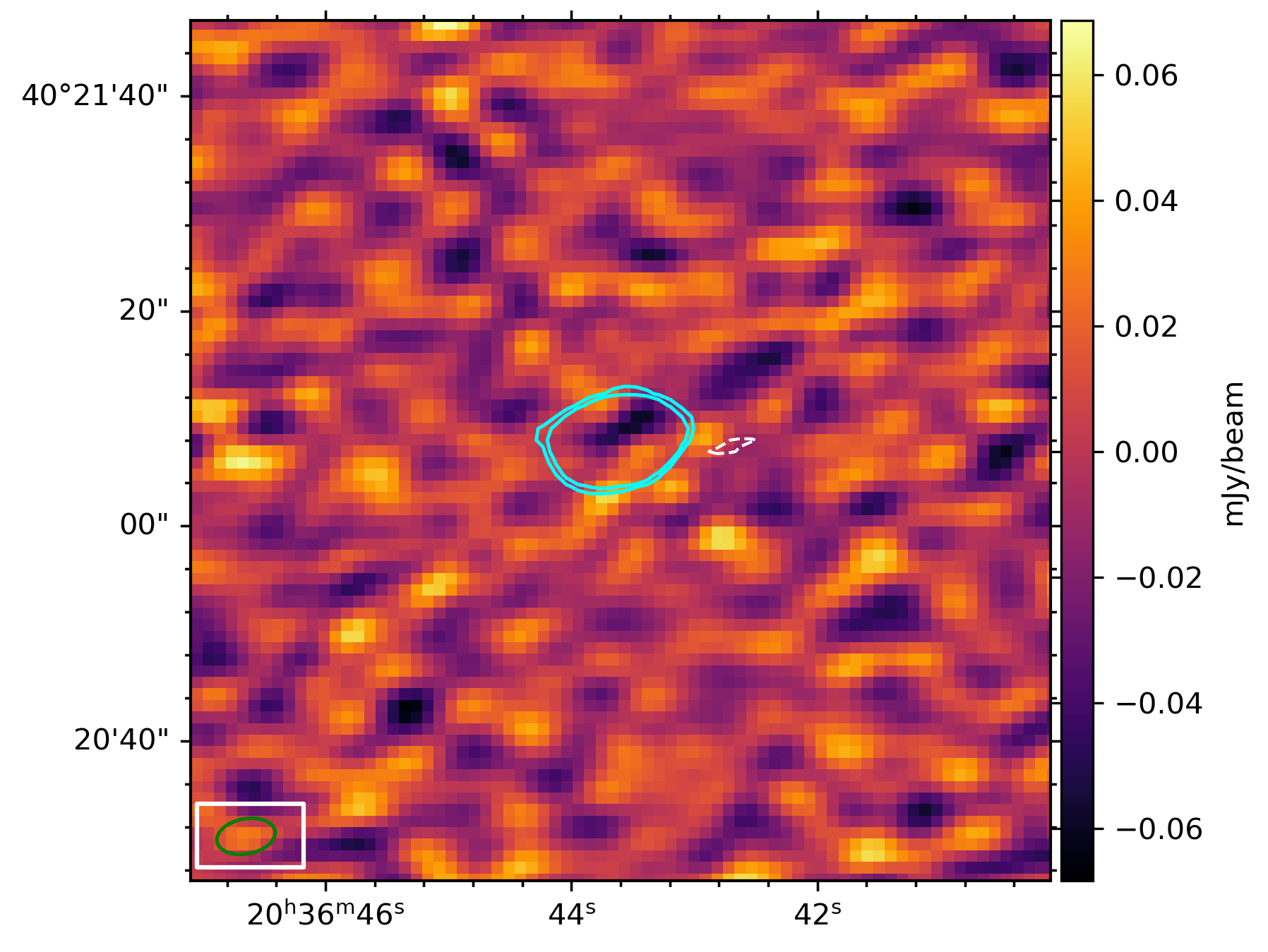} &
\includegraphics[width=0.33\textwidth]{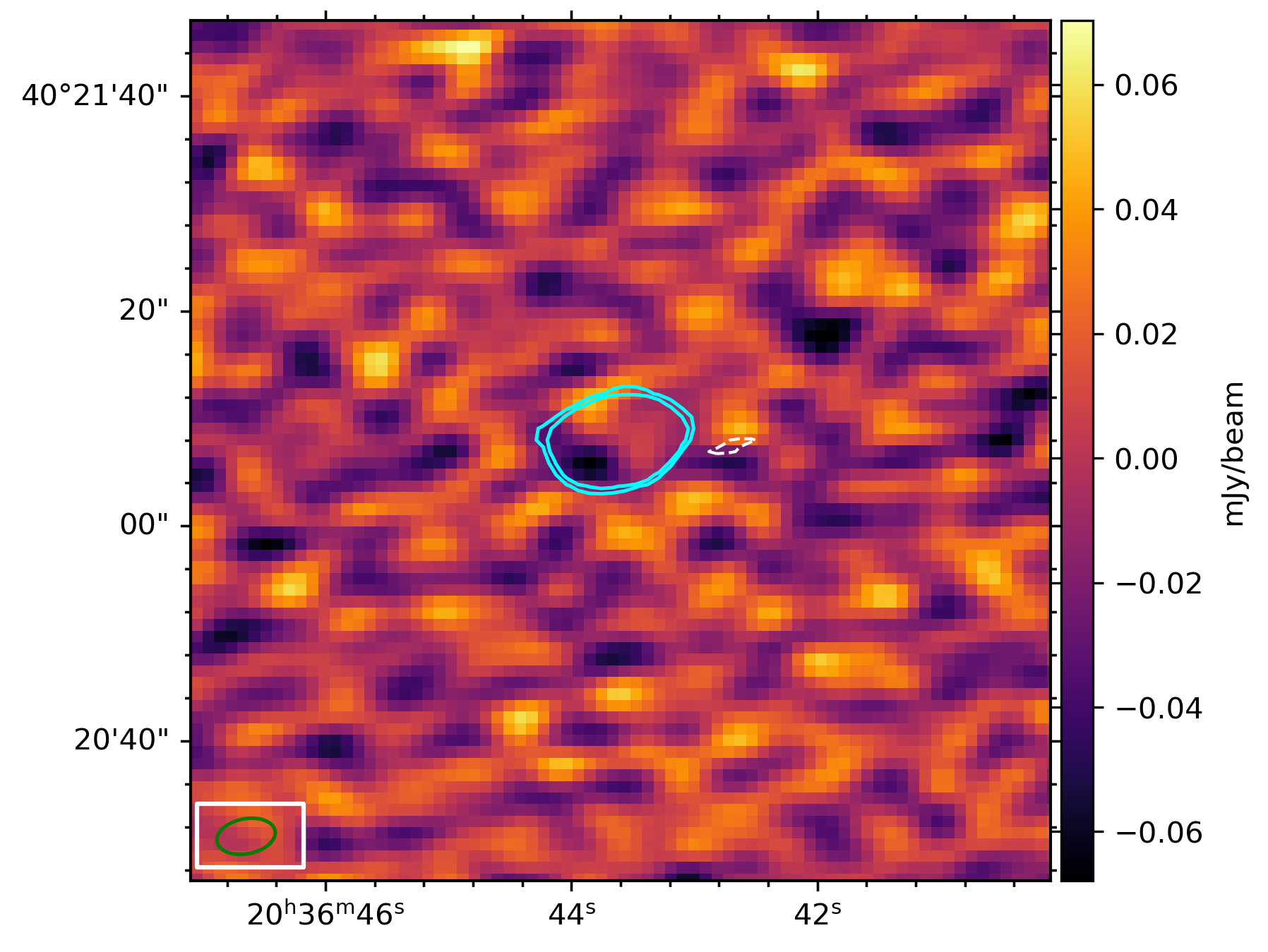} \\

\includegraphics[width=0.34\textwidth]{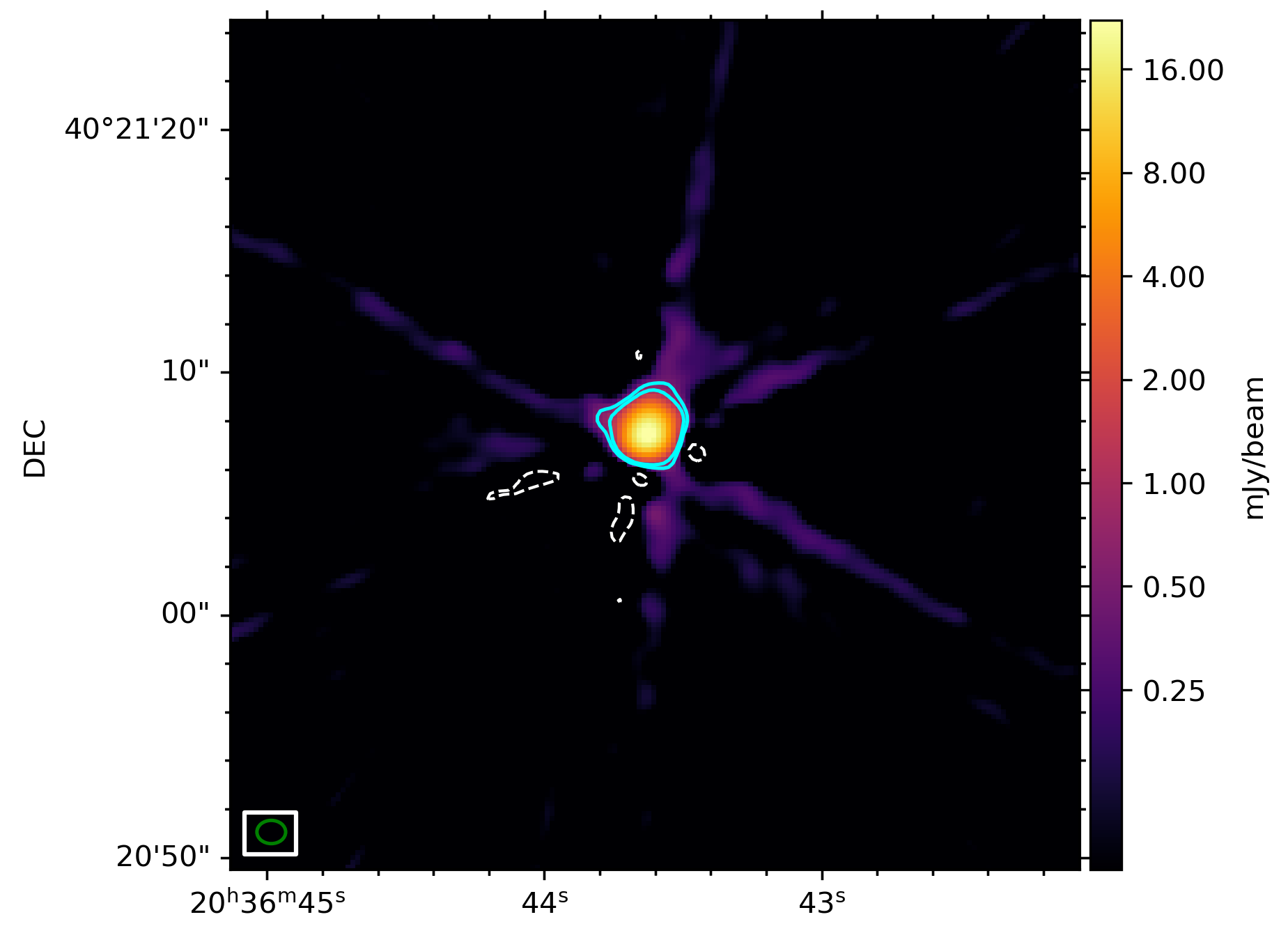} &
\includegraphics[width=0.33\textwidth]{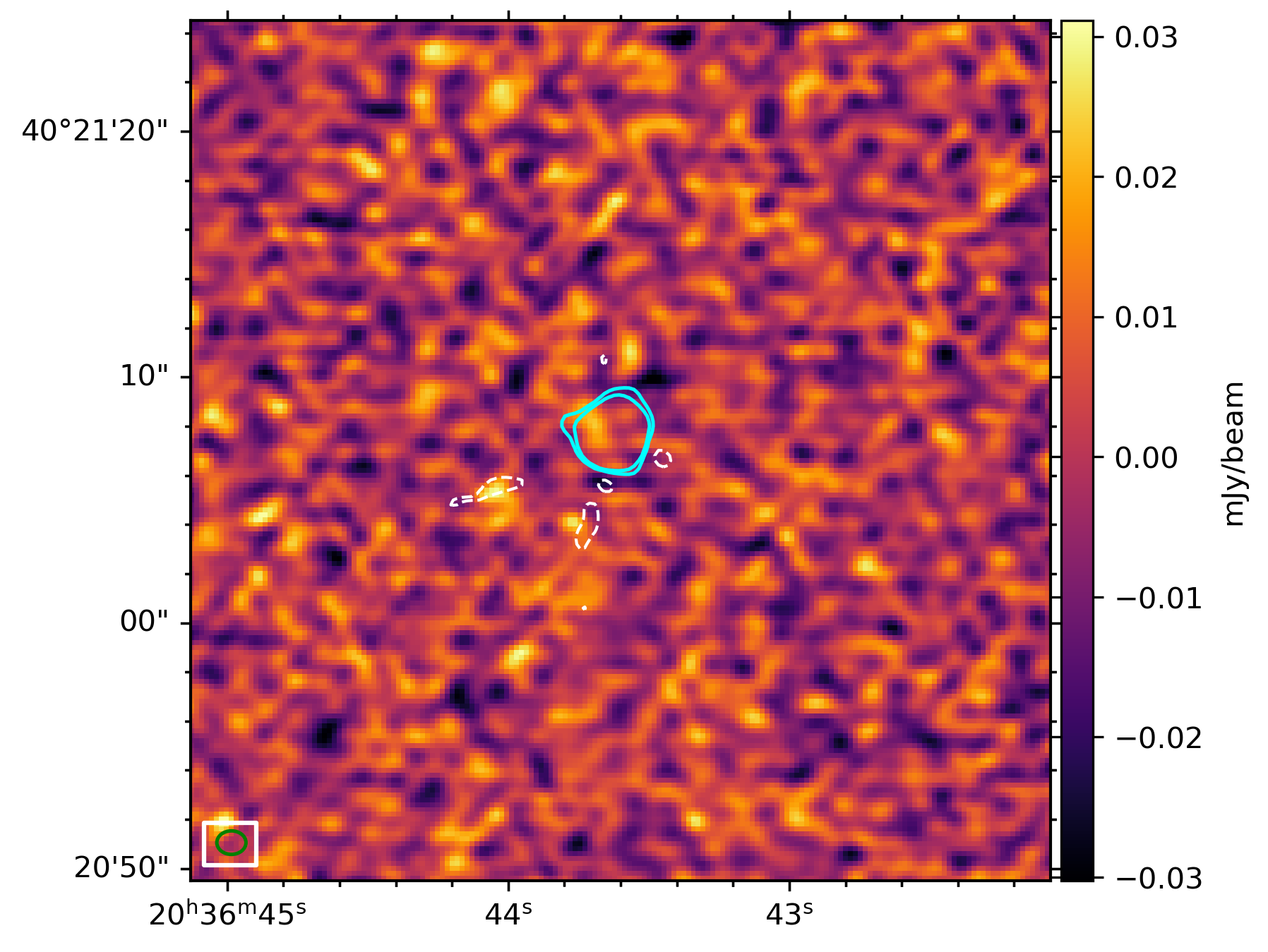} &
\includegraphics[width=0.33\textwidth]{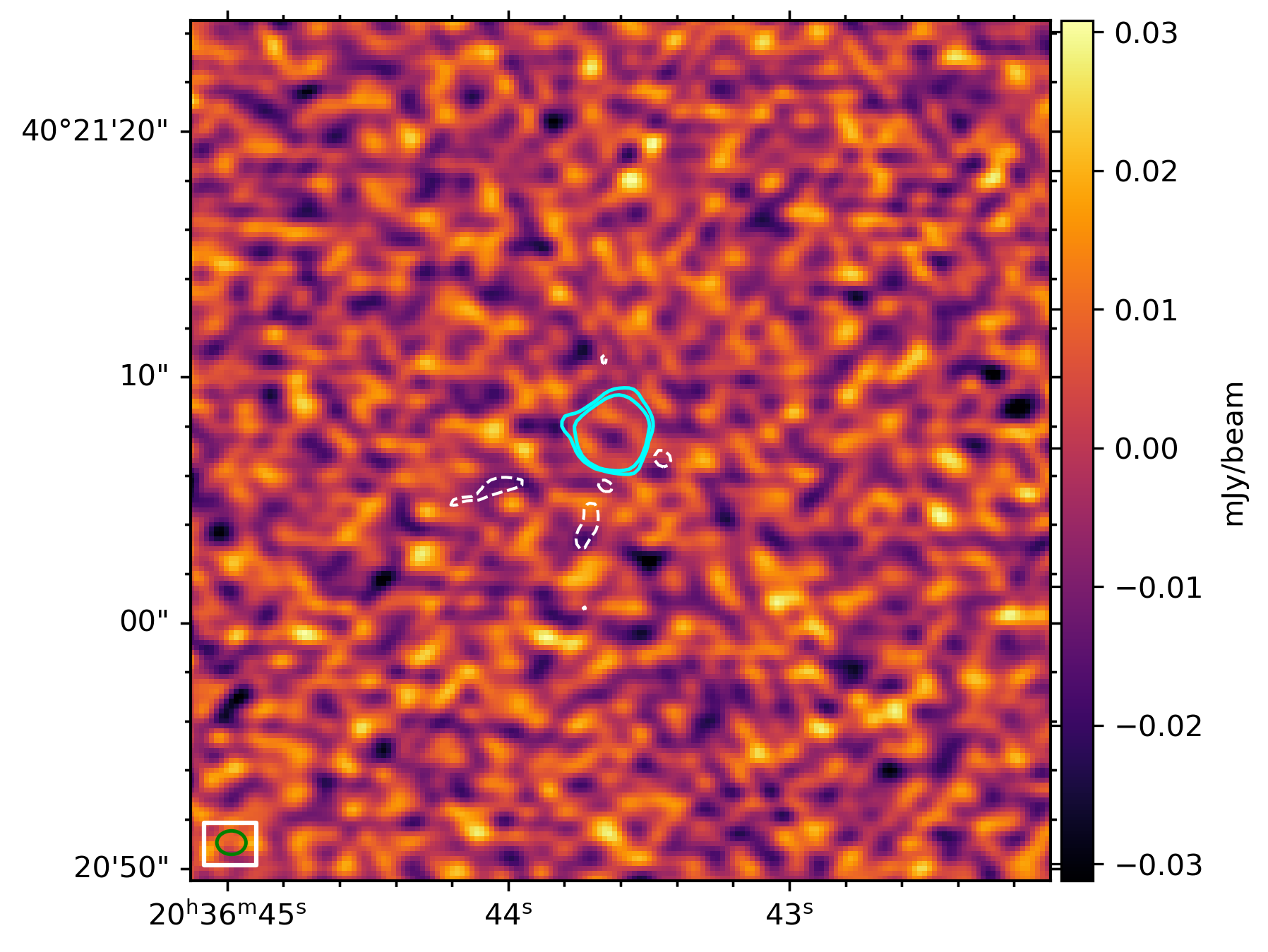} \\

\includegraphics[width=0.34\textwidth]{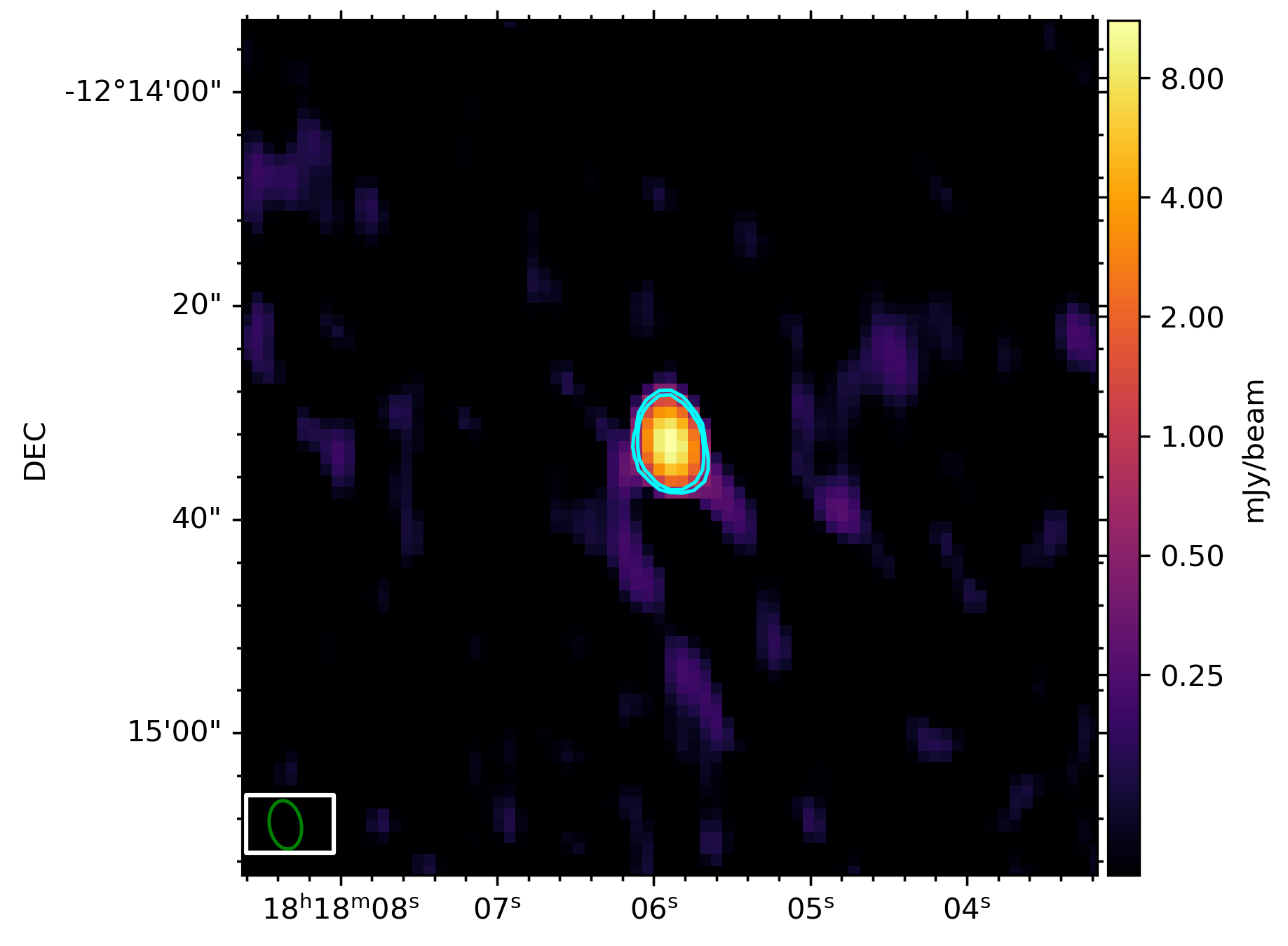} &
\includegraphics[width=0.33\textwidth]{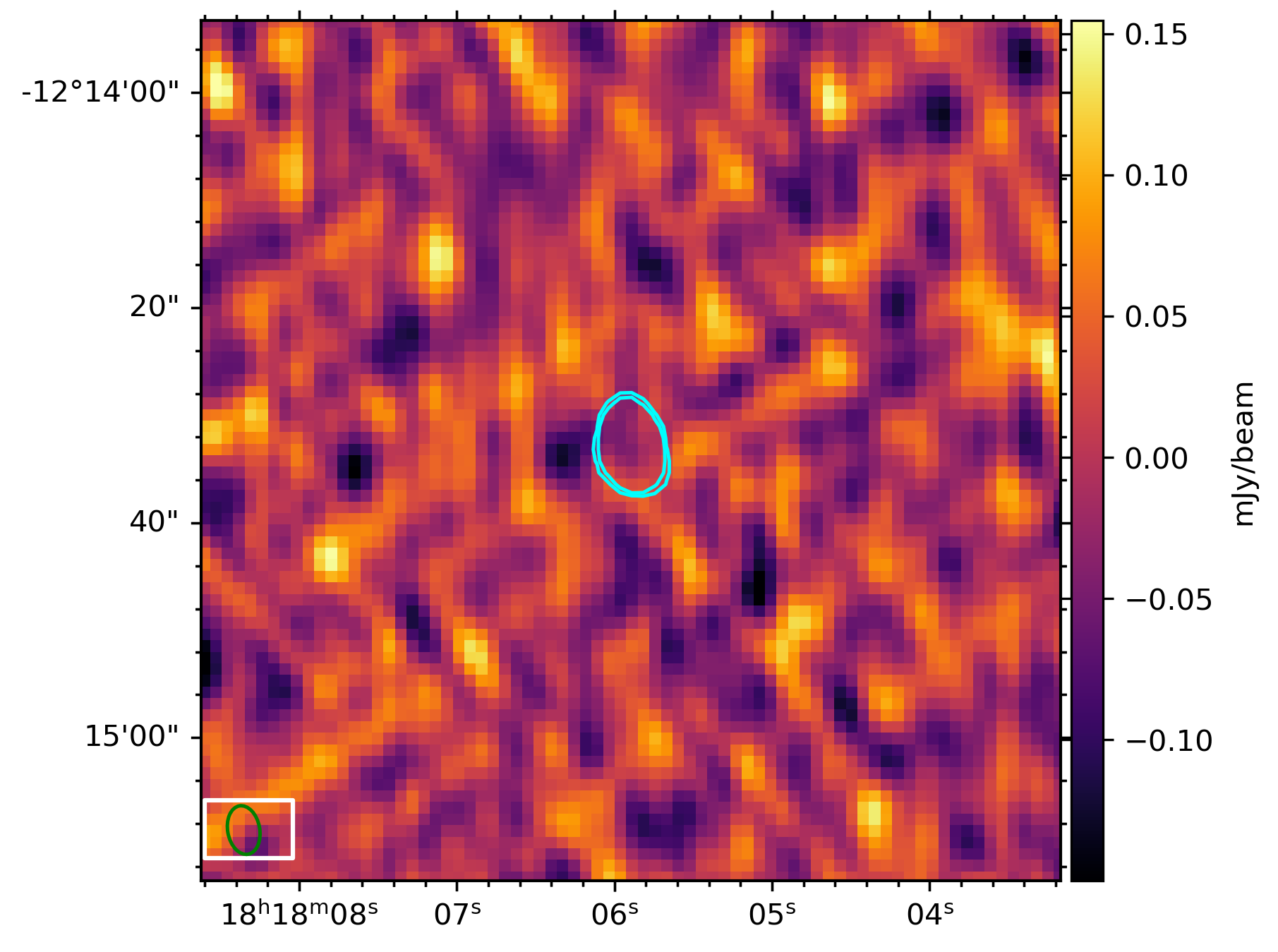} &
\includegraphics[width=0.33\textwidth]{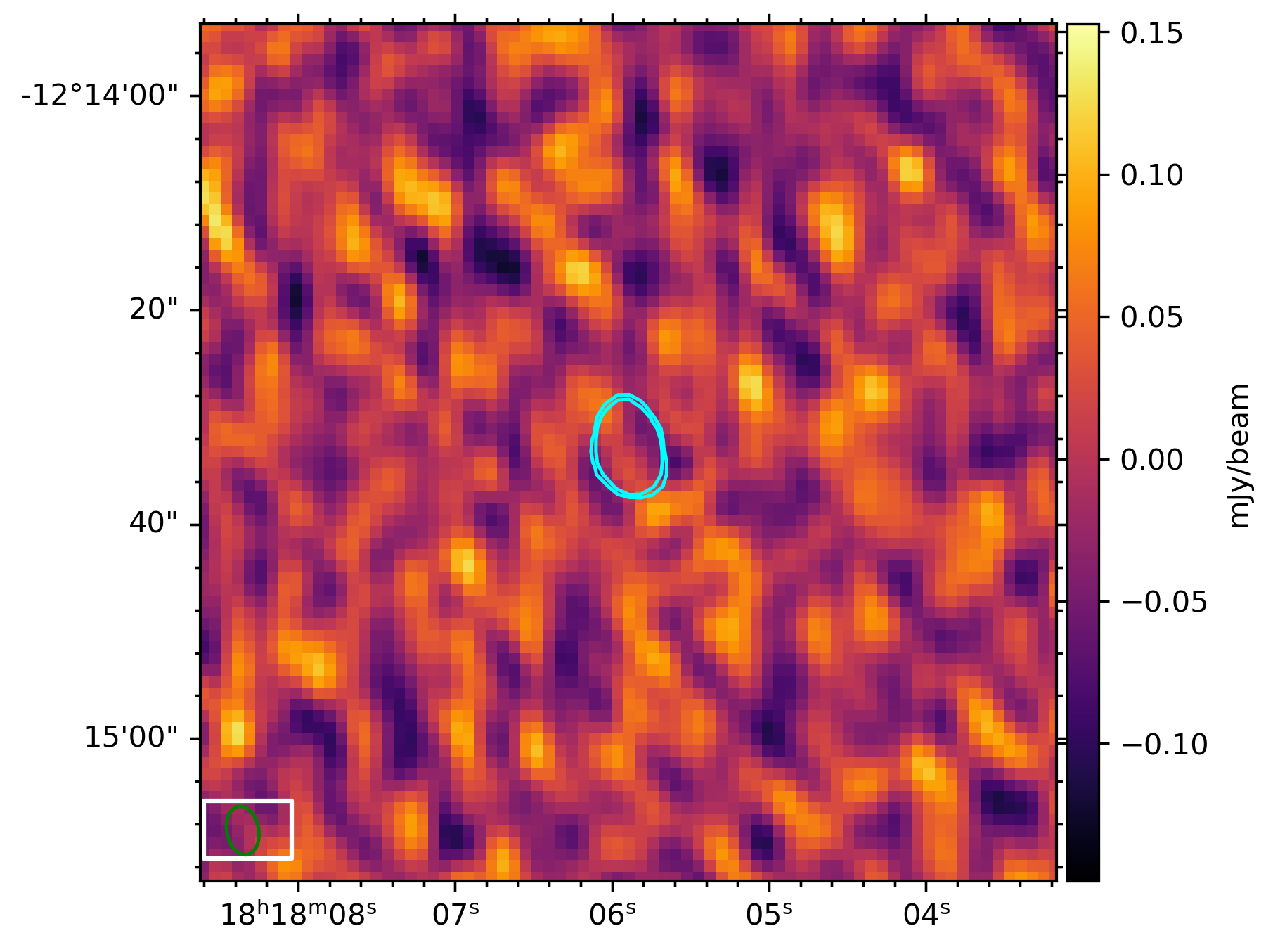} \\

\includegraphics[width=0.34\textwidth]{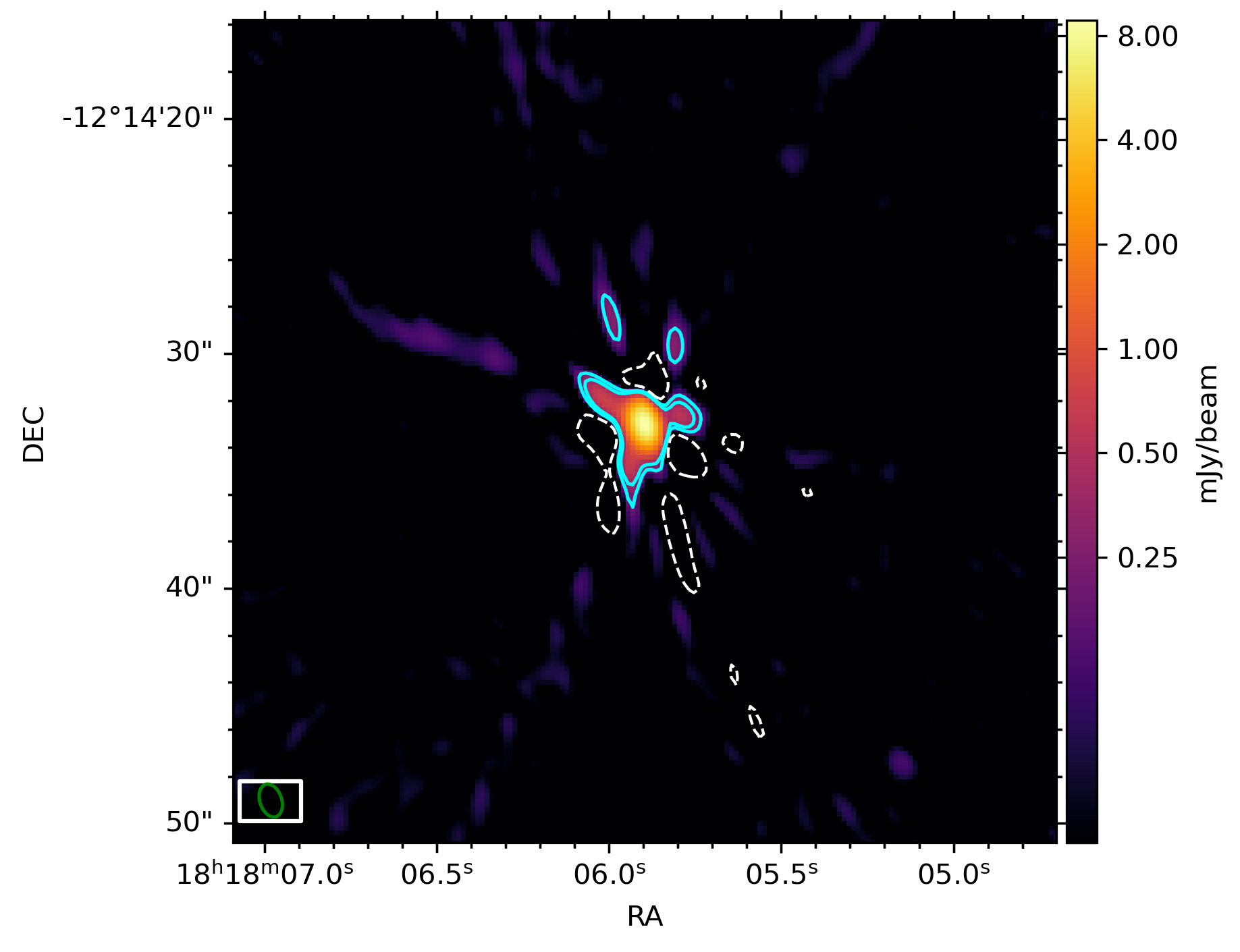} &
\includegraphics[width=0.33\textwidth]{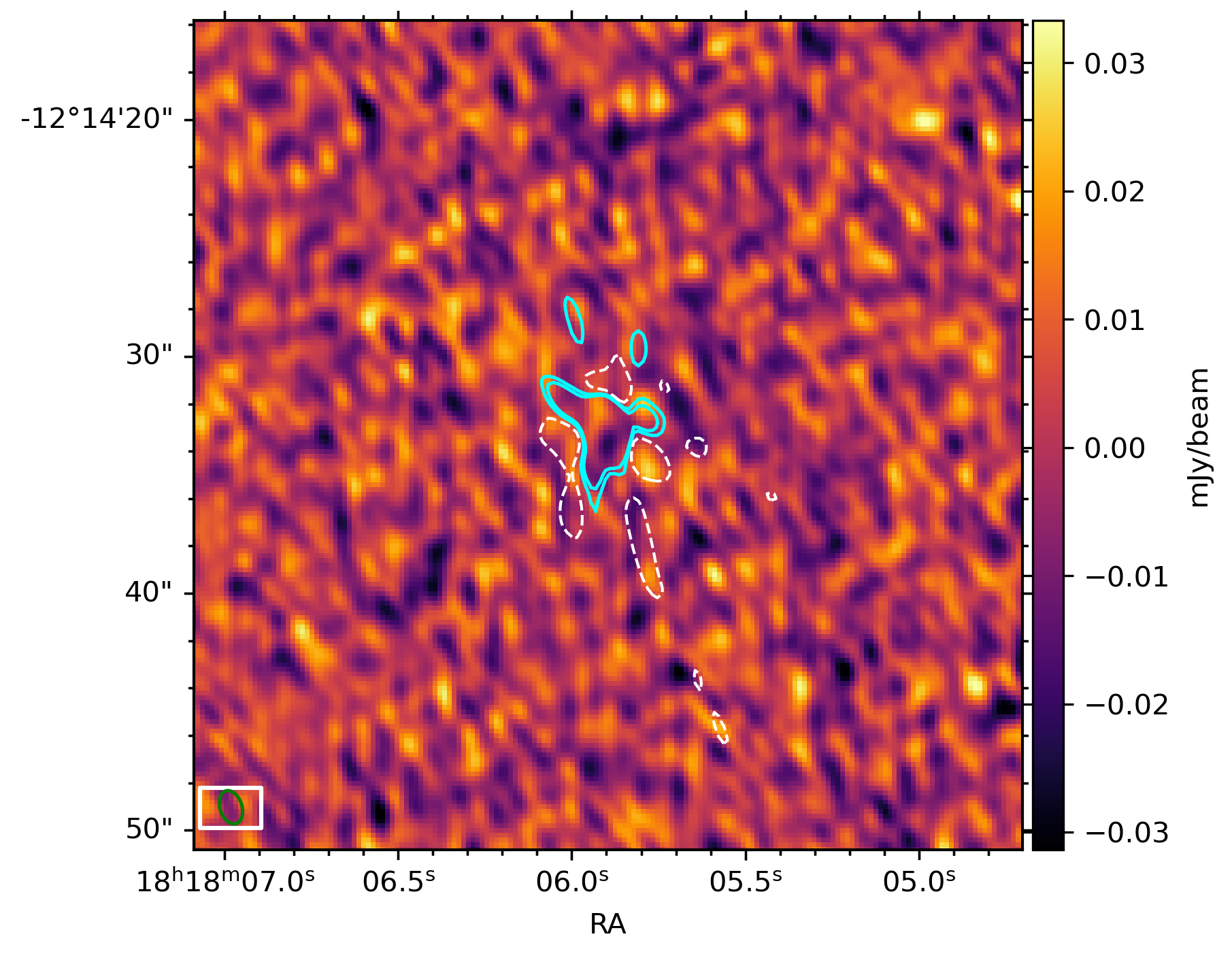} &
\includegraphics[width=0.33\textwidth]{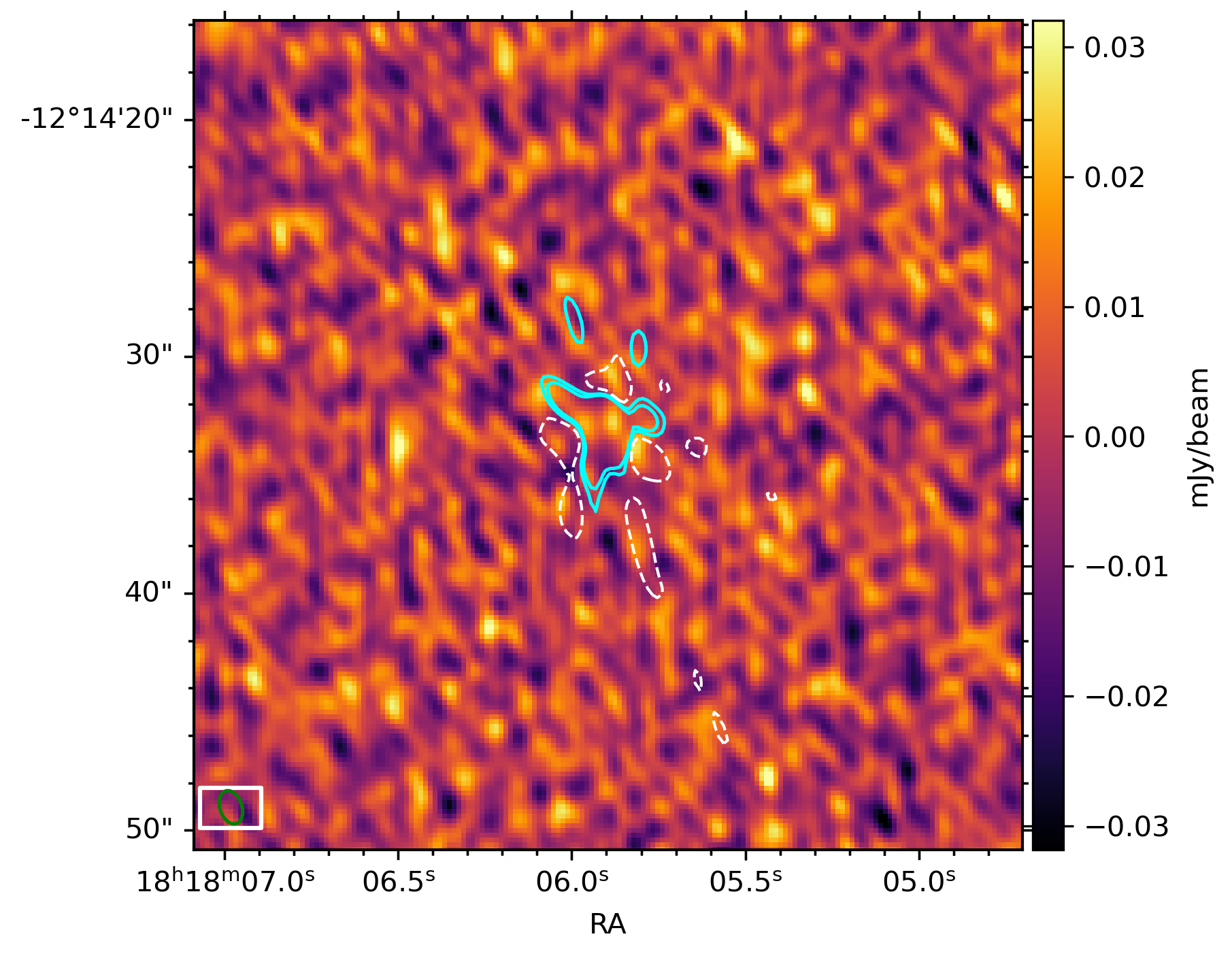} \\

\end{tabular}

\caption{Stokes $I$, $Q$, and $U$ maps (from left to right) of WR~147 and HD~167971.
From top to bottom: WR~147 in the L and C bands, followed by HD~167971 in the L and C bands. The Stokes $I$ maps are shown with a logarithmic color scale. $Q$ and $U$ maps are displayed in a linear autoscale. In all panels, cyan contours at 5 and 9$\sigma$ indicate the target position, while white dashed contours show emission at $-3\sigma$. The synthesized beam is displayed in green in the bottom-left corner.}
\label{Stokes-maps}
\end{figure*}

We detected the systems WR~147 and HD~167971 in the intensity maps obtained in both bands, L and C, as unresolved point sources. However, we did not detect any emission in either band at the position of the target systems in the Stokes maps $Q$ and $U$, where the signature of linear polarization (if present) should be located. The Stokes maps $I, Q,$ and $U$ for WR~147 and HD~167971 in both frequency bands are presented in Fig.~\ref{Stokes-maps}.

To test whether bandwidth depolarization due to Faraday rotation could be affecting our results, we also produced additional Stokes $I$, $Q$, and $U$ maps using narrower frequency ranges in the higher-frequency end of the C band for both WR~147 and HD~167971. As discussed in Sect.\,\ref{bandpassdepol}, given the dependence of Faraday-rotation-driven depolarization processes on wavelength, we restricted this complementary analysis to the C band.

We generated a set of maps using the eight highest-frequency spectral windows of the C-band data (7–8~GHz) and another set using only the two highest-frequency spectral windows (7.75–8~GHz), corresponding to a total bandwidth of 250~MHz. The resulting images do not reveal any detectable polarized emission in either case.  

We obtained flux density measurements for both targets in the L and C bands, together with the rms noise levels of the maps around the source. We present these values, with the synthesized beam of the cleaned Stokes $I$ maps, in Table~\ref{StokesI-Details}. Furthermore, in the same table, we report the values of angular resolution, {\it rms}, and flux density obtained for the narrow, higher-frequency-range images also produced. 

The linearly polarized intensity is defined as $P = \sqrt{Q^{2} + U^{2}}$ and the polarization degree as $\Pi = P/I$. Given the absence of a significant polarization detection, we estimated upper limits on $\Pi$ using the noise properties of the $Q$ and $U$ maps. The rms noise levels, $\sigma_Q$ and $\sigma_U$, were measured over a region coincident with the spatial extent of the source in Stokes $I$. We adopted a conservative approach by using the larger of the two as a representative noise level $\sigma$. The noise in polarized intensity is therefore on the order of $\sigma$, and assuming a confidence level of $3\sigma$, we derive an upper limit on the polarized intensity of $P_{\mathrm{UL}} = 3\sigma$. The corresponding upper limit on the polarization degree is then given by $\Pi_{\rm UL}$ = $P_{\mathrm{UL}} / I$. To ensure consistency in units, the peak Stokes $I$ intensity measured within the same region (in Jy~beam$^{-1}$) was used in the denominator. We report all quantities involved in these estimates, as well as the upper limit on the fractional polarization expressed in \% ($\Pi_\mathrm{UL}^\%$), in Table~\ref{polarization_UL}.

The main result is the total absence of measurable linear polarization for both targets, regardless of the selected spectral band. Our most conservative upper limits on the polarization degree are on the order of 1\,\%.

\begin{table*}[t]

\centering
\fontsize{8.5}{10}\selectfont
\caption{Parameters of the Stokes $I$ maps for both targets.}
\label{StokesI-Details}
\setlength{\tabcolsep}{2.3pt}

\begin{threeparttable}
\begin{tabular}{lccc ccc ccc ccc}
\hline
\noalign{\smallskip}

 & \multicolumn{12}{c}{Observing band} \\
\cline{2-13}
\noalign{\smallskip}

Source
& \multicolumn{3}{c}{L band (1--2 GHz)}
& \multicolumn{3}{c}{C band (4--8 GHz)}
& \multicolumn{3}{c}{C band (7--8 GHz)}
& \multicolumn{3}{c}{C band (7.75--8 GHz)} \\

\noalign{\smallskip}
\cline{2-4}\cline{5-7}\cline{8-10}\cline{11-13}
\noalign{\smallskip}

 & Beam size & {\it rms} & $\mathrm{S_\nu}$
 & Beam size & {\it rms} & $\mathrm{S_\nu}$
 & Beam size & {\it rms} & $\mathrm{S_\nu}$
 & Beam size & {\it rms} & $\mathrm{S_\nu}$ \\

 & ($\arcsec \times \arcsec$) & ($\mu$Jy~b$^{-1}$) & (mJy)
 & ($\arcsec \times \arcsec$) & ($\mu$Jy~b$^{-1}$) & (mJy)
 & ($\arcsec \times \arcsec$) & ($\mu$Jy~b$^{-1}$) & (mJy)
 & ($\arcsec \times \arcsec$) & ($\mu$Jy~b$^{-1}$) & (mJy) \\

\noalign{\smallskip}
\hline
\noalign{\smallskip}

WR~147
& $5.57 \times 3.28$ & 47.6 & $25.5 \pm 0.3$
& $1.20 \times 0.95$ & 21.7 & $38.1 \pm 0.7$
& $0.94 \times 0.76$ & 202.5 & $39.2 \pm 2.3$
& $0.89 \times 0.74$ & 220.8 & $39.1 \pm 2.4$ \\

HD~167971
& $4.61 \times 2.96$ & 86.1 & $12.8 \pm 0.2$
& $1.45 \times 0.94$ & 62.7 & $9.5 \pm 0.4$
& $1.12 \times 0.73$ & 42.3 & $8.3 \pm 0.3$
& $1.07 \times 0.73$ & 68.2 & $8.0 \pm 0.3$ \\

\noalign{\smallskip}
\hline

\end{tabular}

\begin{tablenotes}
\footnotesize
\item Clean beam sizes are given in arcsec,  rms noise levels off-source in
$\mu$Jy~beam$^{-1}$, and flux densities in mJy.
\end{tablenotes}
\end{threeparttable}

\end{table*}

\begin{table}[t]
\small
\centering
\caption{Upper limit on the fractional polarization and the quantities used to derive it.}
\label{polarization_UL}

\setlength{\tabcolsep}{3pt}

\begin{threeparttable}

\renewcommand{\arraystretch}{1.2}
\begin{tabular}{@{}p{4.7cm}ccccc@{}}
\hline
Source 
& $\sigma_Q$ 
& $\sigma_U$ 
& $\sigma$ 
& $I_{\rm peak}$  
& $\Pi_\mathrm{UL}^\%$ \\
\hline
WR~147 -- L band & 26.6 & 26.3 & 26.6 & 22.3 & 0.36 \\
WR~147 -- C band &  9.7 & 6.5 & 9.7 & 24.7 & 0.12 \\
WR~147 -- C band (7--8~GHz) & 26.9 & 15.8 & 26.9 & 24.3 & 0.33 \\
WR~147 -- C band (7.75--8~GHz) & 85.4 & 111 & 111 & 24.0 & 1.38 \\
HD~167971 -- L band & 28.3 & 29.8 & 29.8 & 12.4 & 0.72  \\
HD~167971 -- C band & 10.3 & 11.7 & 11.7 & 9.0 & 0.39  \\
HD~167971 -- C band (7--8~GHz) & 20.7 & 15.9 & 20.7 & 7.9 & 0.79 \\
HD~167971 -- C band (7.75--8~GHz) & 38.4 & 21.7 & 38.4 & 7.6 & 1.52 \\
\hline
\end{tabular}

\begin{tablenotes}
\footnotesize
\item {\it Rms} noise levels ($\sigma_Q$, $\sigma_U$, $\sigma$) are given in
$\mu$Jy~beam$^{-1}$ and peak Stokes $I$ in mJy~beam$^{-1}$.
$\Pi_{\rm UL}^{\%}$ is expressed in \%.
\end{tablenotes}
\end{threeparttable}

\end{table}

\section{Discussion}\label{disc}

The lack of significant linear polarization in the synchrotron sources studied deserves some discussion. Despite the a priori expectation of a polarization signature, the specific physical conditions that prevail in these objects have to be considered to relocate our results in their appropriate framework.

\subsection{Maximum polarization degree}\label{maxpd}
Assuming a fully ordered magnetic field, the maximum intrinsic polarization degree ($\Pi_\mathrm{max}$) of a pure synchrotron source is determined based on the emissivities ($j_\perp$ and $j_\parallel$) for both polarization directions, namely the directions perpendicular and parallel to the projected direction of the magnetic field on the sky plane, respectively. It can be expressed as a function of the electron index ($p$) of the relativistic electron population \citep{RL1979}, namely,
\begin{equation}
 \Pi_\mathrm{max} = \frac{j_\perp - j_\parallel}{j_\perp + j_\parallel} = \frac{p + 1}{p + \frac{7}{3}}.    
\end{equation}

Expressing $\Pi_\mathrm{max}$ as a function of the optically thin synchrotron spectral index ($\alpha_\mathrm{NT} = (p - 1)/2$), we obtain
\begin{equation}
 \Pi_\mathrm{max} = \frac{3\,(\alpha_\mathrm{NT} + 1)}{3\,\alpha_\mathrm{NT} + 5}.    
\end{equation}

For WR~147, \citet{Tasseroul2025} obtained $\alpha_\mathrm{NT} = 0.7$. This leads to $\Pi_\mathrm{max} = 0.72$. In the case of HD~167971, our flux density measurements reported in Table\,\ref{StokesI-Details} (L and 7.75--8\,GHz bands) give a spectral index of about 0.3, leading to $\Pi_\mathrm{max} = 0.66$. These numbers confirm the rather weak dependence of $\Pi_\mathrm{max}$ on $\alpha_\mathrm{NT}$, with values that are always on the order of 0.7.

We clarify that the synchrotron spectra of these sources are optically thin with respect to synchrotron self-absorption. The drop in $\Pi$ expected for synchrotron self-absorbed sources is thus not applicable to our targets of interest. Because FFA is not sensitive to the polarization direction, there is no reason to consider that the alteration of the spectrum by FFA affects the polarization degree. The values derived above are considered as the maximum possible values for the degree of polarization, which would be valid for a fully ordered magnetic field, without considering thermal dilution and other depolarization processes that are addressed below.

\subsection{Thermal dilution}\label{tdilution}
When addressing the question of the polarization of a composite radio source (made of spatially unresolved thermal and NT components), it is relevant to estimate the fraction of the radio flux that is actually of synchrotron origin. The latter component is indeed the only one that has the potential to reveal a polarization signature.

For WR~147, an important insight comes from the detailed analysis of the radio spectral energy distribution performed by \citet{Tasseroul2025}. In their Fig.\,4, the best-fit composite model allows us to evaluate the contribution of synchrotron origin at any frequency. The synchrotron contribution, accounting for attenuation by internal FFA, is expressed as
\begin{equation}\label{eq:SEDiFFA}
S_\nu^\mathrm{NT} = S_{\nu,0}^\mathrm{NT}\,\nu^{-\alpha_\mathrm{NT} + \beta}\,\left[1 - e^{-(\nu/\nu_\mathrm{i-FFA})^{-\beta}}\right],
\end{equation}
\noindent where $S_{\nu,0}^\mathrm{NT}$ is a normalization factor, $\alpha_\mathrm{NT}$ is the NT index of the optically thin synchrotron spectrum, $\beta$ is the index of the internal-FFA absorption coefficient, and $\nu_\mathrm{i-FFA}$ is the internal-FFA turnover frequency. We refer to \citet{Tasseroul2025} for details. Using best-fit parameters ($S_{\nu,0}$ = 18.90 $\pm$ 0.24~mJy\,GHz$^{-\alpha_\mathrm{NT}}$; $\beta = 0.89 \pm 0.03$; $\nu_\mathrm{i-FFA} = 5.20 \pm 0.70$ GHz) and a fixed value for $\alpha_\mathrm{NT} (= 0.7)$, we estimated the NT contribution to the composite radio emission.

Let us define the thermal dilution factor ($f_\mathrm{dil}$) as the ratio of the synchrotron flux density to the total flux density, including the thermal contribution,
\begin{equation}
f_\mathrm{dil} = \frac{S_{\nu}^\mathrm{NT}}{S_{\nu}^\mathrm{NT} + S_{\nu}^\mathrm{T}},
\end{equation}
\noindent where $S_{\nu}^\mathrm{T} = S_{\nu,0}^\mathrm{T}\,\nu^{\alpha_\mathrm{T}}$. In agreement with the formalism used by \citet{Tasseroul2025}, the thermal contribution is defined by its normalization factor ($S_{\nu,0}^\mathrm{T}$) and a thermal spectral index ($\alpha_\mathrm{T}$). While the former was found to be equal to $4.52 \pm 0.10$\,mJy\,GHz$^{-\alpha_\mathrm{T}}$, the latter was fixed to a value of 0.75. This value is significantly steeper than the canonical index predicted by \citet{WB1975} and is more typical of the thermal radio emission from WR stellar winds \citep{Nugis1998}.

At the central frequencies of the L and C bands (1.5 \,GHz and 6.0 \,GHz, respectively), the best-fit model of the radio spectral energy distribution of WR~147 allows us to determine $f_\mathrm{dil,L} = 0.76 \pm 0.01$ and $f_\mathrm{dil,C} = 0.47 \pm 0.02$. As a result, the maximum expected polarization degree is expected to be lowered by that factor. The thermal dilution effect is especially strong for WR~147 because of its unusual bright thermal emission component. Taking into account thermal dilution, the maximum expected polarization degree for WR~147 becomes $\Pi_\mathrm{dil} = f_\mathrm{dil}\,\Pi_\mathrm{max} \approx 0.55$ and 0.34, respectively, for the L and C bands.

Alternatively, for HD~167971, the thermal emission is much weaker. At a distance on the order of 1.7~kpc, the thermal emission at these frequencies should be well below the 0.1 mJy level \citep{DeBecker2018, DeBecker2024}. The thermal emission components of the O-type winds in the system are therefore very small compared to the synchrotron emission, resulting in $f_\mathrm{dil} \approx 1$ in both bands. Thus, the expected maximum value $\Pi$ is not affected.

\subsection{Bandpass depolarization}\label{bandpassdepol}

To consolidate the lack of polarization signature and discard the possible influence of Faraday rotation in our results, we replicated the analysis in two narrower, high-frequency sub-bands, as stated in Sect.~\ref{results}. 

When integrating over a broad bandwidth, Faraday rotation causes the linear polarization angle to vary across the band, as a result of its $\lambda^2$-dependence. This inevitably leads to a significant loss of polarization due to the averaging of differently rotated polarization vectors. This effect is expected to be reduced at higher frequencies (i.e., shorter wavelengths) and when using narrower frequency intervals (where there is less spread and/or scatter of the angle from the lower to the upper boundary of the sub-band). This is why we completed our analysis by producing Stokes maps in narrower bands, especially using the higher frequency spectral windows in the C band, as stated in Sect.\,\ref{results}. 

The absence of polarization detection even in these cases consolidates the lack of a measurable polarization signature at a level greater than about 1\%. As a result, the lack of linear polarization is not attributable to the sole bandpass depolarization effect. The same conclusion applies to the case of WR\,146 investigated by \citet{Hales2017}, where mitigating the effect of bandpass depolarization following the same approach did not lead to the detection of any linear polarization signature. Other depolarization effects appear to be active as addressed in Sect.\,\ref{nopolar}.

\subsection{Lack of synchrotron polarization in PACWBs}\label{nopolar}

The effective polarization degree that is measured is not only dependent on the thermal dilution. The magnetic field in the synchrotron emission region is not perfectly ordered. According to \citet{Burn1966}, the reduction in the polarization degree due to the nonordered magnetic field is expressed by the ratio of energy densities in the ordered and total (i.e., ordered+turbulent) magnetic fields (the so-called turbulent dilution). In addition, wavelength-dependent depolarization effects (based on Faraday rotation) are known to further decrease the polarization degree. As a result, we may express the observed polarization degree as
\begin{equation}\label{depolar}
\Pi_\mathrm{obs} = \Pi_\mathrm{dil}\,\frac{B_0^2}{B_0^2 + B_\mathrm{turb}^2}\,D_\lambda,
\end{equation}
\noindent where $B_0$ and $B_\mathrm{turb}$ are ordered and turbulent magnetic field strengths, and $D_\lambda$ is the depolarization factor (function of wavelength $\lambda$) due to propagation effects. The latter consists typically of Faraday rotation with an amplitude that is dependent on the distribution of the rotation measure (RM) along the line of sight.

The amplitude of the impact of turbulent dilution is directly related to the turbulent-to-total ratio of magnetic energy densities. In the synchrotron emission region of PACWBs, we can only speculate on this ratio. In the very likely scenario where DSA is responsible for particle acceleration, a nonzero turbulent component is absolutely required. A purely uniform magnetic field would compromise the iterative back-scattering on turbulent scattering centers required for significant energy gain of charged particles involved in the particle acceleration process. Thus, turbulent dilution is very likely to significantly reduce the effective polarization fraction.

The important role of depolarization processes in the lack of polarization signature measured for WR~146 has been well documented by \citet{Hales2017}. Depolarization processes are reviewed by \citet{Burn1966}. These include differential Faraday rotation, where synchrotron photons produced at various depths are exposed to different levels of rotation measure. They lead $D_\lambda$ (see Eq.\,\ref{depolar}) to be significantly lower than 1. Given the $\lambda^2$-dependence of Faraday rotation that drives these depolarization processes, one anticipates that this effect is significantly more pronounced in the L band, compared to the C band. Adding to these physical processes, we also have to take into account beam depolarization, where various lines of sight characterized by different rotation measures are averaged in the instrumental beam, which also plays a role. This effect is highly dependent on the size of the synthesized beam. For instance, if a gradient in the magnetic field direction occurs or if there is a distribution of RM across the beam solid angle, the average polarization signature will be reduced. Therefore, this leads to an additional attenuation factor in addition to those accounted for in Eq.\,\ref{depolar}. In systems characterized by complex geometries such as PACWBs, low angular resolution polarimetry is very likely affected by beam depolarization.

\subsection{Synchrotron polarization in a broader context}\label{generalpolar}
In the extragalactic context, radio jets of active galactic nuclei (AGNs) are also well-known synchrotron radio sources. Typical values for the polarization degree are in the range of a few~\%, with some cases reaching a few 10~\% \citep{Hovatta2012, Blinov2021}. Here, again, a nonuniform magnetic field at the scale of the beam size and, to some extent, Faraday depolarization are the main causes for the drop in polarization degree \citep{Yushkov2024}. 

In SNRs, which present similarities to some extent with PACWBs, polarization degrees as low as a few 10$\%$ are interpreted as a result of significant turbulent dilution \citep{Reynolds2012}. Several SNRs seem to show polarization degrees in the range of 20--30\% \citep{Shimoda2022}. However, higher values of the polarization degree have also been measured in some remnants \citep{Dubner2015}.

Protostellar jets are also known to be able to produce synchrotron radiation and are thus worth investigating from the point of view of their linear polarization. The inner lobes of the southern jet associated with the massive protostar IRAS 18162–2048 (specifically HH\,80 and HH\,81) are known to show a linear polarization degree of a few 10\,\% \citet{Carrasco2010, RodriguezKamenetzky2025}. However, \citet{Cheriyan2026} reported on a lack of polarization signature in the knots of HH~80 and HH~81, with upper limits of 1--2\%. This result is similar to the upper limits we derived for WR~147 and HD~167971. The absence of significant polarization was attributed by the authors to turbulence and beam depolarization. In addition, we might also expect some thermal dilution to contribute to the lack of measured polarization, given the nature of the jets that are mainly made of a thermal plasma.

A last example worth mentioning is that of bow shocks produced by runaway massive stars traveling at high peculiar velocities in the interstellar medium. A specific case of interest is EB\,27. The study by \citet{Benaglia2021} provided evidence for NT radio emission,  although no linear polarization was detected. It is interesting to note that for all the examples mentioned above, significant detections of linear polarization arise from resolved sources, where at least the effect of beam depolarization is partly mitigated. However, the case of EB\,27 shows that spatially resolved polarimetric measurements are not necessarily sufficient to warrant the detection of polarization in a synchrotron source. 

When discussing the case of PACWBs in the broader astrophysical landscape, we can clearly see that the usual factors leading to a lowering of the polarization degree, namely, nonuniform magnetic field and Faraday depolarization, are valid for all synchrotron sources. However, PACWBs and, to some extent, jets associated with protostellar objects offer, moreover, ideal conditions for thermal dilution, leading in some cases to a stronger lowering of the maximum expected polarization degree. This effect is not really active for AGNs or SNRs, as such objects do not produce significant thermal radio emission in addition to their synchrotron radiation. In contrast, it becomes important in systems with strong stellar winds capable of generating substantial thermal emission within the spectral range used for the polarimetric analysis. When this thermal component is not spatially resolved from the synchrotron-emitting region, it leads to the thermal dilution of the polarized signal. In addition, because the thermal free-free emission of the winds increases with frequency, this dilution will be stronger at higher radio frequencies (and more so in the C band than in the L band). The typical case is that of PACWBs made up of at least one WR component, whose dense winds can produce a significant thermal contribution, as in the case of WR~146 \citep{Hales2017} and WR~147 (this study). For O-type systems such as HD~167971, thermal dilution is not that important provided that the frequencies explored are not especially high. Thus, the other effects discussed above (i.e., turbulent magnetic field, Faraday-rotation-driven depolarization, beam depolarization) appear to be strong enough to lead to a drastic drop in the polarization signal.

\section{Conclusions}\label{concl}

We present a search for linear polarization in the synchrotron radio emission from two bright particle-accelerating colliding-wind binaries. No linear polarization was detected from either target in either of the observed bands, even after narrowing the bandwidth to attenuate band depolarization effects. These nondetections indicate that the signature of the expected polarization in these systems is heavily suppressed, despite the presence of synchrotron radiation. Our results suggest that no single mechanism is responsible for this; rather, several physical and observational effects act simultaneously to reduce the expected polarization degree to levels that remain below current detection limits.

The lack of a detection of polarization is most likely explained by a combination of a highly turbulent magnetic field in the synchrotron-emitting region (i.e., the colliding-wind region), significant Faraday depolarization, and beam depolarization due to unresolved structure.

In addition, thermal dilution plays an important role in PACWBs, particularly in systems hosting stars with strong thermal winds, such as WR components, where unpolarized free–free emission further contaminates the observable polarization degree. Together, these effects make the detection of linear polarization in such systems intrinsically challenging.

Future progress on this issue will likely require observations that combine a higher angular resolution and improved sensitivity, potentially at higher frequencies where Faraday depolarization is reduced. However, the decreasing synchrotron brightness at such frequencies makes these observations challenging. 
Next-generation facilities, such as the ngVLA and SKA-Mid, or VLBI observations at suitable frequencies, could provide the necessary capabilities to isolate the synchrotron-emitting regions from the thermal winds and potentially reveal polarized emission. Such observations would offer valuable constraints on magnetic field structure and particle acceleration processes in colliding-wind binaries, helping to disentangle intrinsic emission properties from depolarization effects.

\begin{acknowledgements}

This research is part of the PANTERA-Stars collaboration, an initiative aimed at fostering research activities on the topic of particle acceleration associated with stellar sources\footnote{\url{https://www.astro.uliege.be/~debecker/pantera}}. This research has made use of NASA's Astrophysics Data System Bibliographic Services. This publication benefits from the support of the F.R.S.-FNRS (Wallonia-Brussels Federation, Belgium) in the context of the FRIA Doctoral Grant awarded to A. B. Blanco. We want to thank the referee for their careful reading of the paper and for their positive feedback and encouraging words.
     
\end{acknowledgements}

% Please note that we have included the references below in
% order to compile the document, but we ask you to:
\bibliographystyle{aa} % style aa.bst
  \bibliography{biblio.bib} % your references Yourfile.bib

\end{document}